\documentclass[aps,prd,12pt,preprint,tightenlines,superscriptaddress,
amsfonts,amssymb,amsmath,byrevtex,showpacs,nofootinbib]{revtex4-1}
\usepackage{graphics}
\usepackage{graphicx}

\begin{document}
\newcommand{\dR}{\mathbb R}
\newcommand{\dC}{\mathbb C}
\newcommand{\dS}{\mathbb S}
\newcommand{\dZ}{\mathbb Z}
\newcommand{\id}{\mathbb I}
\newcommand{\dM}{\mathbb M}
\newcommand{\dH}{\mathbb H}
\newcommand{\tm}{\tilde{\mu}}
\newcommand{\tn}{\tilde{\nu}}

\title{Reduced phase space approach to Kasner universe\\ and the problem of time in quantum theory}

\author{Przemys{\l}aw Ma{\l}kiewicz
\\
Institute for Gravitation and the Cosmos, Department of Physics, The Pennsylvania State University, University Park, PA 16802, USA.
\\ and
\\ Department of Fundamental Research,
National Centre for Nuclear Research, \\
Ho\.{z}a 69, 00-681 Warsaw, Poland.
\\ pmalk@fuw.edu.pl}

\date{\today}
\begin{abstract}

We apply the reduced phase space quantization to the Kasner universe. We construct the kinematical phase space, find solutions to the Hamilton equations of motion, identify Dirac observables and arrive at physical solutions in terms of
Dirac observables and an internal clock. We obtain the physical Hilbert space, which is the carrier space of the self-adjoint representation of the Dirac observables. Then we discuss the problem of time. 
We demonstrate that the inclusion of evolution in a gravitational
system, at classical level as well as at quantum level, leads respectively
to canonically and unitarily inequivalent theories. The example of Hubble operator in two different clock variables and with two distinct spectra is given.

\end{abstract}

\maketitle

\section{Introduction}
The standard model of cosmology is based on the remarkably simple
solution to general relativity, the
Friedman-Robertson-Walker (FRW) universe. It is expected, however,
that a slightly perturbed FRW universe, when evolved back in time,
at some moment close enough to the big bang singularity, will lose
its space-like symmetries. Therefore, in order to understand
the singular conditions from which the universe emerged nearly 14
billion years ago, a study of more general cosmological spacetimes
is needed.

A general solution of general relativity in the vicinity of
cosmological singularity has been studied by Belinskii,
Khalatnikov and Lifshitz (BKL) in \cite{bkl}. In the BKL scenario,
as spacetime approaches singularity, the time derivatives of
the gravitational field are shown to dominate over all spatial
derivatives for relatively long stretches of time. Surprisingly,
the evolution of the general gravitational field turns out to be well
approximated, at each point separately, by a sequence of the so-called Kasner epochs. Each epoch is a vacuum solution to the
homogenous spacetime model of Bianchi I type. The transitions
between epochs are the effect of non-negligible spatial curvature,
which arises quickly and vanishes after a relatively short period
of time. In the BKL scenario the universe undergoes an infinite
number of chaotic-like transitions and eventually collapses into a
singularity in a finite proper time.

It is commonly believed that the incompleteness of classical
theory, which breaks down at the singularity, will be overcome by
quantization of the gravitational degrees of freedom. For this
purpose, the Dirac method of quantization is usually employed (see e.g. \cite{PAM,WdW,LewAsh}). In this paper we focus on quantum theory of the Kasner universe. We follow, however, an alternative way to
quantum theory, namely the reduced phase space quantization (see e.g. \cite{Mal,Ash}). Since the Kasner model plays a central role
in the BKL description of a generic cosmological singularity, we
believe that the present and future investigations into this model supported by current and forthcoming astrophysical and
cosmological data can help obtaining new insights into the universe's origin.

The fact that in canonical general relativity the evolution of gravitational fields coincides with gauge transformation or equivalently that there is no privileged time standard to measure motion, gives rise to the so-called problem of time (see e.g. Kuchar \cite{Kuch}). The problem of time consists of a few related though distinguishable issues. Following Kuchar's terminology, we will treat in this paper the most fundamental issue, namely {\it the multiple choice problem}. In essence, it states that two different choices of time may produce different quantum theories. We will show how severe the problem is and examine its origin. In the view of the results obtained in this paper, the proposals for explaining the multiple choice problem existing in the literature are unsatisfactory. For example, in Isham \cite{Ish} we can read that two different choices of time lead to canonically equivalent theories, which admit unitarily inequivalent quantum representations due to the Van-Hove phenomenon \cite{vH}. We will show that in fact the multiple choice problem can be traced back to the canonical formulation of general relativity and thus studied at classical level.

The application of the reduced phase space approach to the
Kasner universe turns out to
be manageable and quite straightforward. The system consists of a single constraint on the
kinematical phase space, which is six-dimensional. We identify the (physical)
reduced phase space, that is the space of Dirac observables, which is
four-dimensional. We consider two examples of {\it clock variables} to introduce the {\it physical} evolution and
construct the so-called true Hamiltonians. We
argue that the evolution is free of the singularity present in the
classical theory. We compare the spectra of the Hubble operator in two different clock variables and show that they are very different.

Before we start let us introduce the notation that will be used
throughout the text. The canonical variables, which follow
from the Legendre mapping applied to the Einstein-Hilbert action,
parametrize the kinematical phase space, denoted by $\mathcal{P}$. In
this space, the Hamiltonian constraint $H$ is introduced. The
constraint surface, defined by $H=0$, is denoted by $\mathcal{S}$.
The Dirac observables, denoted by $\mathcal{D}_i$, are defined by
the relation $\{\mathcal{D}_i,H\}=0$ and with the domain
restricted to the constraint surface. The space of Dirac
observables, called the reduced phase space, or the physical phase
space, will be denoted by $\mathcal{P}_R$. In Rovelli \cite{rovel}, it is proposed to call kinematical phase space functions by
partial observables in order to emphasize that they can be {\it
measured} by observers, though the outcome of such a measurement
made alone cannot be {\it predicted} by theory. There is, however,
one kinematical degree of freedom for which the theory makes the
immediate prediction, i.e. the Hamiltonian constraint vanishes, $H=0$.
In what follows, by partial observables we mean functions
restricted to $\mathcal{S}$ (which seems to be slightly different from
Rovelli's notion), and which will be denoted by $\mathcal{P}_i$.

\section{Lagrangian formulation} The
Hilbert-Einstein action reads
\begin{equation}\label{he}
\mathcal{S}_{HE}=\int_{\Omega\subset \mathcal{M}} R\sqrt{-g}~d^3xdt
\end{equation}
where $g,R$ are the metric determinant and the Ricci scalar,
respectively. The integral is taken over an open subset $\Omega$
of the manifold $\mathcal{M}$. We specify action (\ref{he}) to the
case of vacuum Bianchi I model with $\mathcal{M}=R\times\Sigma$,
where $\Sigma$ is a compact spacelike leaf and we assume the
following metric type:
\begin{equation}\label{metric}
ds^2=-N^2(t)dt^2+a_1^2(t)(dx^1)^2+a_2^2(t)(dx^2)^2+a_3^2(t)(dx^3)^2
\end{equation}
which leads (\ref{he}) to the form (see Appendix \ref{calc})
\begin{equation}\label{k}
\mathcal{S}_{HE}=2\int_{t_0}^{t_1}dt\int_{\Sigma}Na_1a_2a_3
\bigg[\sum_i\frac{1}{Na_i}\bigg(\frac{\dot{a}_i}{N}\bigg)_{,t}+\sum_{i>j}\bigg(\frac{\dot{a}_i}{Na_i}\bigg)\bigg(\frac{\dot{a}_j}{Na_j}\bigg)\bigg]d^3x
\end{equation}
Applying the variational principle to (\ref{k}) gives the Lagrange
equations:
\begin{equation}\label{le}
\sum_{i>j}\frac{\dot{a}_i}{a_i}\frac{\dot{a}_j}{a_j}=0~,~~\dot{a}_i\dot{a}_j=N\bigg(\frac{\dot{(a_ia_j)}}{N}\bigg)_{,t}~\forall_{i\neq
j}
\end{equation}
The solutions were first found by Kasner in \cite{kas}.
Later, they were rediscovered by Taub in \cite{taub}, who
gave the solutions in the following form:
\begin{equation}\label{taub}
ds^2=-dt^2+t^{2p_1}dx^2+t^{2p_2}dy^2+t^{2p_3}dz^2
\end{equation}
where the constants $p_1$, $p_2$ and $p_3$ satisfy:
\begin{equation}
p_1+p_2+p_3=1,~~p_1^2+p_2^2+p_3^2=1
\end{equation}
All the above solutions, except for $p_1=1$, $p_2=1$ or $p_3=1$,
admit a {\it cosmological singularity} for $t=0$.

It is easily seen that action (\ref{k}), equations of motion
(\ref{le}) and solution (\ref{taub}) are invariant under any time
re-parameterization such that $t\mapsto \tilde{t}(t)$ and
$N\mapsto\frac{dt}{d\tilde{t}}N$. This is the only gauge freedom,
which is preserved under the reduction of the Hilbert-Einstein
action to the homogenous spacetime. Therefore, by {\it gauge
transformation} we mean any change of time parameter $t\mapsto
\tilde{t}(t)$.

\section{Hamiltonian formulation} Using action (\ref{k})
we define the momenta\footnote{For the sake of simplicity, from
now on we drop the integration over the compact space-like leaf
$\Sigma$ whenever it should appear.}:
\begin{equation}\label{leg}
\pi^{N}:=\frac{\partial L}{\partial
\dot{N}}=0~,~~\pi^{i}:=\frac{\partial L}{\partial
\dot{a}_i}=-\frac{2}{N}\dot{(a_ja_k)}
\end{equation}
The Legendre mapping (\ref{leg}) is singular, and its range is the
submanifold of the phase space given by $\pi^{N}=0$. The
Hamiltonian reads:
\begin{equation}
H_0=\pi^{N}\dot{N}+\sum\pi^{i}\dot{a}_i-L=\frac{1}{2}\sum_{i}\pi^i\dot{a}_i
\end{equation}
The Dirac analysis \cite{PAM} leads to the reduction of phase
space by the conjugate pair $(N,\pi^{N})$ and the introduction of
the Hamiltonian constraint:
\begin{equation}\label{con}
H:=\frac{N}{8}\sum_{i\neq j\neq
k}\frac{\pi^i}{a_ja_k}\big(\pi^ia_i-\pi^ja_j-\pi^ka_k)\approx 0
\end{equation}
where now the lapse $N$ is a Lagrange multiplier. One may verify that the
vanishing of Hamiltonian constraint (\ref{con}) and Hamilton's equations are
equivalent to the first and the second of the Euler-Lagrange
equations in (\ref{le}), respectively.
\subsection{New variables and motion} We introduce the new canonical
variables:
\begin{equation}\label{newleg}
X_i:=\frac{1}{2}a_ja_k~,~~P_i:=-4\frac{\dot{a}_i}{N}
\end{equation}
in which the Hamiltonian constraint (\ref{con}) reads:
\begin{equation}\label{con2}
H=-\frac{N}{4\sqrt{2}}\sum_{i>j}\sqrt{\frac{X_iX_j}{X_k}}P_iP_j
\end{equation}
The symplectic form on the kinematical phase space is defined as:
\begin{equation}\label{sym}
\omega:=\sum_i dX_i\wedge dP_i
\end{equation}
and its minus inverse is the Poisson bracket, i.e.:
\begin{equation}
-\omega^{-1}=\sum_i\bigg(\frac{\partial}{\partial
X_i}\frac{\partial}{\partial P_i}-\frac{\partial}{\partial
P_i}\frac{\partial}{\partial X_i}\bigg)
\end{equation}
The Hamilton equations in the gauge $N=4\sqrt{2}\sqrt{X_1X_2X_3}$
read:
\begin{eqnarray}\label{heq1}
\dot{X}_i=\frac{\partial H}{\partial P_i}=-X_i(X_jP_j+X_kP_k)\\
\label{heq2}\dot{P}_i=-\frac{\partial H}{\partial
X_i}=P_i(X_jP_j+X_kP_k)
\end{eqnarray}
From combining the above equations into one:
\begin{equation}
\dot{X}_iP_i+X_i\dot{P}_i=(X_iP_i)_{,t}=0
\end{equation}
we obtain that $\Gamma_i:=X_iP_i$ are constants of motion (as we will
see in a moment they may be identified with some of the Dirac
observables). Putting this back to (\ref{heq1}), (\ref{heq2}) we
find that:
\begin{equation}\label{sol}
X_i(t)=X_i(t_0)e^{-(\Gamma_j+\Gamma_k)(t-t_0)}~,~~P_i(t)=P_i(t_0)e^{(\Gamma_j+\Gamma_k)(t-t_0)}
\end{equation}
where $P_i(t_0)$ and $X_i(t_0)$ are the initial conditions for the
Hamilton equations and $t_0$ will be specified later. The physical
solutions (\ref{sol}) should satisfy the Hamiltonian constraint
(\ref{con2}), which can be rewritten now as:
\begin{equation}\label{con3}
\sum_{i>j}\Gamma_i\Gamma_j\approx 0
\end{equation}
The change of the arrow of time $t\mapsto -t$ in (\ref{sol}) is
equivalent to the sign change $\Gamma_i\mapsto -\Gamma_i$ for all
$i$'s, so we can add the condition:
\begin{equation}\label{arrow}
\sum_i\Gamma_i>0
\end{equation}
which ensures that the singularity is approached as the time $t$
grows. Moreover, the following three cases:
\begin{equation}\label{milne}
\Gamma_1=\Gamma_2=0,~~\Gamma_2=\Gamma_3=0,~~\Gamma_1=\Gamma_3=0
\end{equation}
can be shown to correspond to the
Milne space, which can be isometrically embedded in Minkowski
spacetime and thus are non-singular (i.e. the coordinates are
singular, not the spacetime itself). We exclude them from the phase
space.

\subsection{Dirac's observables}

It is known that for the Hamiltonian satisfying $dH\neq 0$ in a
neighborhood of the constraint surface $H=0$, one may locally
introduce such a canonical parametrization of the kinematical
phase space that the canonical coordinates:
\begin{equation}
(X_i,P_i),~~i=1,\dots,n
\end{equation}
are replaced with the new canonical pairs:
\begin{equation}
(H,T), (\tilde{X}_i,\tilde{P}_i),~~i=1,\dots,n-1
\end{equation}
such that the variable $T$ is canonically conjugate to $H$ and the
symplectic form now reads:
\begin{equation}
\omega=dT\wedge dH+d\tilde{X}_i\wedge d\tilde{P}_i
\end{equation}
It is now easily seen that the space of functions which commute
with the Hamiltonian $H$ is given by:
\begin{equation}
\{H,\mathcal{D}_i\}=0\Rightarrow\mathcal{D}_i=\mathcal{D}_i(\tilde{X}_i,\tilde{P}_i, H)
\end{equation}
which restricted to the constraint surface $H=0$ can be identified with the following space of functions:
\begin{equation}
\mathcal{D}_i=\mathcal{D}_i(\tilde{X}_i,\tilde{P}_i)
\end{equation}
An easy way to find all the Dirac observables is by pulling back the
symplectic form $\omega$ to the constraint surface:
\begin{equation}
\omega|_{H=0}=d\tilde{X}_i\wedge d\tilde{P}_i
\end{equation}
and ensuring the pulled-back two-form is in canonical form. Obviously, this recipe does not
depend on gauge as
\begin{equation}
\omega=(NH)\cdot dT\wedge d\frac{1}{N}+\frac{1}{N}\cdot dT\wedge
d(NH)+d\tilde{X}_i\wedge d\tilde{P}_i
\end{equation}
so $\omega|_{NH=0}=\omega|_{H=0}$ for any $N\neq 0$.

In what follows, we will obtain the complete set of Dirac observables by
restricting the symplectic form $\omega$ introduced in (\ref{sym})
to the constraint surface (\ref{con3}). Applying the mapping:
\begin{eqnarray}\nonumber
\Gamma_1&=&\frac{1}{2}x-\frac{1}{2}y+z\\
\Gamma_2&=&\frac{1}{2}x-\frac{1}{2}y-z\\ \nonumber
\Gamma_3&=&\frac{3}{4}x+\frac{5}{4}y
\end{eqnarray}
one arrives at the following form of the constraint (\ref{con3}):
\begin{equation}\label{cone}
x^2-y^2-z^2=0
\end{equation}
The constraint is solved in the new coordinates $(r,\phi)\in
R\times S$ such that:
\begin{equation}
x=r~,~~y=r\cos\phi~,~~z=r\sin\phi
\end{equation}
so that $\Gamma_i$'s read:
\begin{eqnarray}
\Gamma_1&=&\frac{1}{2}r(1-\cos\phi+2\sin\phi)\\
\Gamma_2&=&\frac{1}{2}r(1-\cos\phi-2\sin\phi)\\
\Gamma_3&=&\frac{1}{4}r(3+5\cos\phi)
\end{eqnarray}
One checks that setting the arrow of time in (\ref{arrow}) gives
\begin{equation}
\sum_i\Gamma_i=\frac{r}{4}(7+\cos\phi)>0
\end{equation}
and leads to $r>0$. The exclusion of the Milne space cases
(\ref{milne}), which are equivalent to $\phi= 0$ and $\cos\phi=
-\frac{3}{5}$, restricts the parameter $\phi$ in the following
way:
\begin{equation}
\phi\in (0,\phi_1)\cup(\phi_1,\phi_2)\cup(\phi_2,2\pi)
\end{equation}
where $\phi_1$ and $\phi_2$ are the solutions to
$\cos\phi=-\frac{3}{5}$. The starting kinematical phase space was
parameterized by the six coordinates $(X_i,P_i)$. Since $X_i>0$, we
could use the coordinates $(X_i,\Gamma_i)$ as well. Thus,
constraint surface (\ref{con3}) may be parameterized by the five
coordinates $(X_i,r,\phi)$.

The complete set of Dirac observables and their commutation
relations can be found by restricting the symplectic form $\omega$
in $\mathcal{P}$ to the constraint surface $\mathcal{S}$. Let us
denote the embedding of the constraint surface in the kinematical
phase space by $E:\mathcal{S}\mapsto\mathcal{P}$ so that
$E^*:\mathcal{C}^{\infty}(\mathcal{P})\ni (P^i,X_i)\mapsto (r,\phi,X_i)\in
\mathcal{C}^{\infty}(\mathcal{S})$ is the
restriction of the kinematical phase space functions to the
constraint surface functions:

\begin{align}
\notag &E^*\big(\sum dX_i\wedge dP_i\big)=E^*\big(\sum dX_i\wedge
\frac{d\Gamma_i}{X_i}\big)\\ \notag
&=d\big[(\frac{1}{2}-\frac{1}{2}\cos\phi+\sin\phi)\ln
X_1+(\frac{1}{2}-\frac{1}{2}\cos\phi-\sin\phi)\ln
X_2+(\frac{3}{4}+\frac{5}{4}\cos\phi)\ln X_3\big]\wedge dr\\
\notag &+d\big[(\frac{1}{2}r\sin\phi+r\cos\phi)\ln
X_1+(\frac{1}{2}r\sin\phi-r\cos\phi)\ln
X_2-\frac{5}{4}r\sin\phi\ln X_3\big]\wedge d\phi\\
&=d\mathcal{O}_1\wedge dr+d\mathcal{O}_2\wedge d\phi
\end{align}
where we have defined:
\begin{eqnarray}
\mathcal{O}_1&=&\big(\frac{1}{2}-\frac{1}{2}\cos\phi+\sin\phi\big)\ln
X_1+\big(\frac{1}{2}-\frac{1}{2}\cos\phi-\sin\phi\big)\ln
X_2+\big(\frac{3}{4}+\frac{5}{4}\cos\phi\big)\ln X_3\\
\mathcal{O}_2&=&\big(\frac{1}{2}r\sin\phi+r\cos\phi\big)\ln
X_1+\big(\frac{1}{2}r\sin\phi-r\cos\phi\big)\ln
X_2-\frac{5}{4}r\sin\phi\ln X_3
\end{eqnarray}

The space of Dirac's observables is called the reduced phase space
$\mathcal{P}_R: (r,\phi,\mathcal{O}_1,\mathcal{O}_2)\in R_+\times
I_1\cup I_2\cup I_3\times R\times R$. Apparently, the space of
Dirac observables in the Kasner universe is not simply connected.
The four Dirac observables together with any constraint surface
function $t$ such that
\begin{equation}
\{t,H\}\big|_{H=0}\neq 0
\end{equation}
form a coordinate system $(r,\phi,\mathcal{O}_1,\mathcal{O}_2,
t)$ on the five-dimensional constraint surface $\mathcal{S}$
equipped with the two-form:
\begin{equation}\label{romega}
\omega_S:=\sum dX_i\wedge dP_i\bigg|_{H=0}=d\mathcal{O}_1\wedge
dr+d\mathcal{O}_2\wedge d\phi.
\end{equation}
induced from the symplectic form $\omega$ on the kinematical phase
space $\mathcal{P}$. It should be added that the freedom in the
choice of the fifth coordinate $t$ on $\mathcal{S}$ is bigger then
the freedom in choosing the lapse function $N:\mathcal{S}\mapsto R_+$, which fixes only the first derivative of $t$ with respect to the Hamiltonian vector field, i.e. $\{t,H\}\big|_{H=0}=N^{-1}$, where $H$ itself is taken with the lapse equal 1.

\subsection{Partial observables}

Solutions found in (\ref{sol}) include both the physical and
non-physical sector. In the physical sector, however, all the
solutions should be expressible in terms of coordinates on
$\mathcal{S}$. Setting $X_1(t_0)=1$, we express all the constants
occurring in (\ref{sol}), that is $\Gamma_i, X_i(t_0), P_i(t_0)$,
in terms of the Dirac observables
$\mathcal{O}_1,\mathcal{O}_2,\phi,r$ and arrive at:
\begin{eqnarray}
P_1&=&\frac{1}{2}r(1-\cos\phi+2\sin\phi)\exp(\frac{1}{4}r(5+3\cos\phi-4\sin\phi)(t-t_0))\\
P_2&=&\frac{1}{2}r(1-\cos\phi-2\sin\phi)\exp(\frac{1}{4}r(5+3\cos\phi+4\sin\phi)(t-t_0))e^{-\frac{5\mathcal{O}_1\sin\phi+\frac{1}{r}\mathcal{O}_2(3+\cos\phi)}{-5-3\cos\phi+4\sin\phi}}\\
P_3&=&\frac{1}{4}r(3+5\cos\phi)\exp(r(1-\cos\phi)(t-t_0))e^{-\frac{\mathcal{O}_1(2\sin\phi-4\cos\phi)+\frac{1}{r}\mathcal{O}_2(-2+2\cos\phi+4\sin\phi)}{-5-3\cos\phi+4\sin\phi}}\\
X_1&=&\exp(-\frac{1}{4}r(5+3\cos\phi-4\sin\phi)(t-t_0))\\
X_2&=&\exp(-\frac{1}{4}r(5+3\cos\phi+4\sin\phi)(t-t_0))e^{\frac{5\mathcal{O}_1\sin\phi+\frac{1}{r}\mathcal{O}_2(3+\cos\phi)}{-5-3\cos\phi+4\sin\phi}}\\
X_3&=&\exp(-r(1-\cos\phi)(t-t_0))e^{\frac{\mathcal{O}_1(2\sin\phi-4\cos\phi)+\frac{1}{r}\mathcal{O}_2(-2+2\cos\phi+4\sin\phi)}{-5-3\cos\phi+4\sin\phi}}
\end{eqnarray}
The above solutions are solutions (\ref{sol}) restricted to the
constraint surface $\mathcal{S}$. The variable $t$ parameterizes
the gauge orbits in the constraint surface $\mathcal{S}$. After
the specification of the value of $t_0$ for each gauge orbit, that
is defining $t_0$ as a function of Dirac observables, the variable
$t$ becomes the fifth coordinate, which assigns a specific value
to each point in $\mathcal{S}$.

In analogy to the Friedman cosmology, we will be interested in the
Hubble and deceleration parameters, which in the Kasner model are
introduced for each of the three directions (see Appendix
\ref{calc}) and read:
\begin{eqnarray}\label{h1}
H_1&=&
-\frac{r(1-\cos\phi+2\sin\phi)}{32}\exp\bigg(\frac{r}{4}(7+\cos\phi)(t-t_0)\bigg)
e^{\frac{\mathcal{O}_1(7\sin\phi-4\cos\phi)+\frac{1}{r}\mathcal{O}_2(1+3\cos\phi+4\sin\phi)}{10+6\cos\phi-8\sin\phi}}\\
H_2&=&
-\frac{r(1-\cos\phi-2\sin\phi)}{32}\exp\bigg(\frac{r}{4}(7+\cos\phi)(t-t_0)\bigg)
e^{\frac{\mathcal{O}_1(7\sin\phi-4\cos\phi)+\frac{1}{r}\mathcal{O}_2(1+3\cos\phi+4\sin\phi)}{10+6\cos\phi-8\sin\phi}}\\
H_3&=&
-\frac{r(3+5\cos\phi)}{64}\exp\bigg(\frac{r}{4}(7+\cos\phi)(t-t_0)\bigg)
e^{\frac{\mathcal{O}_1(7\sin\phi-4\cos\phi)\sin\phi+\frac{1}{r}\mathcal{O}_2(1+3\cos\phi+4\sin\phi)}{10+6\cos\phi-8\sin\phi}}\\
\label{q1}
q_1&=&\sqrt{2}~\frac{7+\cos\phi}{1-\cos\phi+2\sin\phi}-1\\
q_2&=&\sqrt{2}~\frac{7+\cos\phi}{1-\cos\phi-2\sin\phi}-1\\
\label{q3} q_3&=&2\sqrt{2}~\frac{7+\cos\phi}{3+5\cos\phi}-1
\end{eqnarray}
We note that the deceleration parameters
$q_i=2\sqrt{2}\big(\frac{\sum\Gamma_j}{\Gamma_i}\big)-1$ are
constants of motion, i.e. Dirac's observables. All the above six
quantities determine and are determined by the components of the connection and
curvature matrices (see Appendix \ref{calc}). Therefore, they
represent the {\it local} properties of the Kasner universe and do not form the complete space of observables in the compact
universe.

\subsection{Gauge transformations}

For $t$ to be a function on $\mathcal{S}$, we need to specify
$t_0$ as a function of Dirac observables. We note that the choice:
\begin{equation}
t_0:=\frac{\mathcal{O}_1(7\sin\phi-4\cos\phi)+\frac{1}{r}\mathcal{O}_2(1+3\cos\phi+4\sin\phi)}{\frac{r}{2}(7+\cos\phi)(5+3\cos\phi-4\sin\phi)}
\end{equation}
simplifies nicely the formulae for the Hubble parameters:
\begin{eqnarray}
H_1&=&
-\frac{r(1-\cos\phi+2\sin\phi)}{32}\exp\bigg(\frac{r}{4}(7+\cos\phi)t\bigg)\\
H_2&=&
-\frac{r(1-\cos\phi-2\sin\phi)}{32}\exp\bigg(\frac{r}{4}(7+\cos\phi)t\bigg)
\\
H_3&=&
-\frac{r(3+5\cos\phi)}{64}\exp\bigg(\frac{r}{4}(7+\cos\phi)t\bigg)
\end{eqnarray}
and leaves the formulas for deceleration parameters unchanged.

We want, however, more than that and apart from simplicity we
require that the time coordinate satisfies two extra conditions:
(a) has a clear physical meaning and (b) the singularity occurs at
its finite value. A distinguished choice is the cosmological time,
for which the lapse function $N_{cos}=1$ and the singularity is
reached at $t_{cos}=0$ (for all gauge orbits).

We use the relation $Ndt=N_{cos}dt_{cos}$ to obtain the formula:
\begin{equation}
t_{cos}=\int_{t_s}^{t}
\frac{dt_{cos}}{dt}~dt=\int_{+\infty}^{t}\frac{N}{N_{cos}}~dt
\end{equation}
where $t=t_s$ defines the four-dimensional boundary of the
constraint surface $\mathcal{S}$, at which the singularity occurs.
Then we insert $N=4\sqrt{2}\sqrt{X_1X_2X_3}$ and obtain
\begin{equation}\label{time}
t_{cos}=\frac{-16\sqrt{2}}{r(7+\cos\phi)}~\exp\bigg(-\frac{r}{4}(7+\cos\phi)t\bigg)
\end{equation}
This time redefinition simplifies the formulae for the Hubble
parameters even further:
\begin{eqnarray}\label{h4}
H_1&=& \frac{1-\cos\phi+2\sin\phi}{\sqrt{2}(7+\cos\phi)~t_{cos}}\\
\label{h5}
H_2&=&\frac{1-\cos\phi-2\sin\phi}{\sqrt{2}(7+\cos\phi)~t_{cos}}
\\ \label{h6}
H_3&=& \frac{3+5\cos\phi}{2\sqrt{2}(7+\cos\phi)~t_{cos}}
\end{eqnarray}
It should be stressed that $t_{cos}$ occurring in the above
formulae is a {\it function} on $\mathcal{S}$. Thus, the
cosmological time $t_{cos}$ may be related to the kinematical
phase space functions. It cannot be done uniquely as there is no natural
projection from the kinematical space to the constraint surface. However, there exists a class of kinematical phase space functions, which coincide with $t_{cos}$ on the constrained surface. To sum up, $t_{cos}$ is a partial observable as much as any $H_i$.

\section{Problem of time}

All internal clocks are given by the formula:
\begin{equation}
\tau=\int_0^{t_{cos}} N^{-1}dt_{cos}+\tau_0
\end{equation}
where $N(t_{cos},\mathcal{D}_i)>0$ is a function of the cosmological time and Dirac
observables and $\tau_0(\mathcal{D}_i)$ is any function of Dirac
observables. Due to the change of time parameter $t\mapsto\tau$,
any time-dependent function on the constraint surface, i.e. any partial
observable, is transformed accordingly:
\begin{equation}
\mathcal{P}(\mathcal{D}_i,t)\mapsto\tilde{\mathcal{P}}(\mathcal{D}_i,\tau)=\mathcal{P}(\mathcal{D}_i,t(\tau,\mathcal{D}_i))
\end{equation}
{\it Thus, in general, $\mathcal{P}$ and $\tilde{\mathcal{P}}$ will
have different dependance on Dirac observables and after
quantization they may have different spectra}. Therefore, it is
meaningless to speak about spectra of partial observables, like
energy density, curvature or volume, without reference to the choice
of internal clock. To support this statement, it is enough to
note that the commutation relations:
\begin{equation}
\{\mathcal{P}_1,\mathcal{P}_2\}\mapsto\{\tilde{\mathcal{P}}_1,\tilde{\mathcal{P}}_2\}=\{\mathcal{P}_1,\mathcal{P}_2\}+
\frac{\partial\mathcal{P}_1}{\partial
t}\{t(\tau,\mathcal{D}_i),\mathcal{P}_2\}+\frac{\partial\mathcal{P}_2}{\partial
t}\{\mathcal{P}_1,t(\tau,\mathcal{D}_i)\}
\end{equation}
becomes altered after the change of time. Note that the commutation relations between the Dirac observables can be obtained from the form $\omega_S$ given in (\ref{romega}), where any variable chosen to play a role of time is interpreted as an external parameter. Thus, in the constraint
surface the gauge transformation is not canonical and consequently
it cannot be unitary in quantum theory. This implicates that the
spectral properties of the same operator in different gauges
should be, in general, different.

The gauge transformation has another interesting feature. Suppose
that we are given any two partial observables, which monotonically
increase with time and whose range is identical. For example, let
it be the curvature $C$ and energy density $\rho$. Then there is
always such a gauge transformation $\tau\mapsto\tau'$ that energy
density in one gauge is functionally identical with the curvature
in another gauge, that is $\rho|_{\tau}\equiv C|_{\tau'}$. {\it In
effect, the spectral properties can be shared by many different
partial observables}. If we drop the assumption about the identity
of ranges, the two different partial observables will still share
the identical dependence on Dirac observables for some time during
the evolution.

Obviously, one is unable to study the cosmological singularity problem
without the notion of evolution, for example just by the
inspection of Dirac observables. {\it Thus, this gives rise to the very
interesting question of the possible dependance between the choice
of time and the fate of singularity in quantum theory}.

\subsection{Geometrical formulation of the problem of time}
Let us take a closer look at the structure of systems
with a Hamiltonian constraint. The constraint equation $H=0$
defines the embedding $E:\mathcal{S}\mapsto\mathcal{P}$ of the
constraint surface $\mathcal{S}$ into the kinematical phase space
$\mathcal{P}$ having its dimension increased by 1. The constraint
surface $\mathcal{S}$ is therefore odd-dimensional. All the
physical motion takes place in the surface and the vectors tangent
to the trajectories (gauge orbits) are given by the
Hamiltonian vector field $X_H$ (gauge generator), defined in
the standard way:
\begin{equation}
X_H:~\omega(\cdot,X_H)=-dH
\end{equation}
where $\omega$ is the symplectic form in $\mathcal{P}$ and the
Hamiltonian vector field $X_H\big|_{H=0}$, which is restricted to the constraint surface, will be
denoted by $X_H$ for brevity.

The constraint surface is equipped with the two-form $\omega_S$
induced from the kinematical phase space, $\omega_S:=E^*\omega$.
The form $\omega_S$ is a singular closed two-form of maximal rank. It
is represented at each point $p\in\mathcal{S}$ by an antisymmetric
matrix $\omega_S(p):T_p\mathcal{S}\mapsto T^*_p\mathcal{S}$ in the tangent space of the odd-dimensional manifold. Its null vector is
$X_H$, satisfying $\omega_S(\cdot,X_H)=0$. The null vector
$X_H$ is a generator of the line bundle with the projection (submersion):
$\pi:\mathcal{S}\mapsto\mathcal{P}_R$ from the constraint surface
to the reduced phase
space. The pullback $\pi^*$ is understood as an injection from the space of functions $\mathcal{D}_i$ such that
$X_H(\mathcal{D}_i)=0$ to the space of all observables. The diagram below illustrates the relations between
$\mathcal{P}$, $\mathcal{S}$ and $\mathcal{P}_R$.

\begin{equation}
\mathcal{P}\xleftarrow{~Embedding, ~E~}\mathcal{S}\xrightarrow{~~
Projection,~\pi~}\mathcal{P}_R
\end{equation}

Since the form $\omega_S$ is singular it cannot be inverted into the Poisson structure (which should be quantized). Suppose we introduce a projection on the tangent space at $p\in\mathcal{S}$, denoted by $P_p:T_p\mathcal{S}\mapsto T_p\mathcal{S}$, such that $P_p(X)=0\Leftrightarrow X\sim X_H(p)$. Then, the matrix $\omega_S(p)$ can be inverted on the restricted domain $\{X\in T_p\mathcal{S}: P_p(X)=X\}$. Though it enables to derive the inverse of $\omega_S$, this construction is ambiguous due to ambiguity in the choice of the projection $P_p$. Now, the (minus) inverse $-\omega_{S,P}^{-1}:T^*_p\mathcal{S}\mapsto T_p\mathcal{S}$ is the Poisson bracket.  

The generalization of the above construction follows straightforwardly. Suppose we attach to each point $q\in\mathcal{S}$ a projection $P_q$ of the considered type. In addition, we assume that there exists a slicing of $\mathcal{S}$ such that the tangent space to a slice at any  $q\in\mathcal{S}$ is identical with the range of $P_q$. Let the slicing be given by a function $t:\mathcal{S}\mapsto R$. Now the Poisson structure for each $q\in\mathcal{S}$ is given by the inverse of the form $\omega_S$ restricted to the hypersurface $t=const$ containing $q$, and let it be denoted by $-\omega_{S,t}^{-1}$.

The construction of $-\omega_{S,t}^{-1}$ may be achieved by exploiting the simple relation that determines the induced Poisson structure:
\begin{equation}
-\omega_{S,t}^{-1}(t,\mathcal{D}_i):=\{t,\mathcal{D}_i\}_{S,t}=0
\end{equation}
with all other commutation relations being fixed uniquely by the independent-of-the-choice-of-time commutation relation between Dirac observables.

\subsection{Canonical transformations in constraint surface}

Once a slicing is introduced in the constraint surface, we arrive at the triple $(\mathcal{S},\omega_{S},t)$. Such a structure is well known in classical mechanics and is called a {\it contact manifold}. The restriction of $\omega_S$ to the constant time hypersurfaces forms a symplectic submanifold with the symplectic form $\omega_{S,t}$. The form $\omega_{S,t}$ can be pulled back to the constraint surface, $\widetilde{\omega}_{S,t}=t^*\omega_{S,t}$. It should be noted that in general $\widetilde{\omega}_{S,t}\neq\omega_S$. The relation between $\widetilde{\omega}_{S,t}$ and ${\omega}_{S}$ is the subject of the theory of canonical transformations and will be discussed below. 

One may think of a canonical transformation as a symplectomorphism, however, the contact manifold provides a better avenue to define this notion. Let us cite the definition of the canonical transformation from Abraham and Marsden \cite{mech}.

{\bf Definition}. Let $(\mathcal{P}_1, \omega_1)$ and
$(\mathcal{P}_2, \omega_2)$ be symplectic manifolds and
$(R\times\mathcal{P}_i, \tilde{\omega}_i)$ the corresponding
contact manifolds. A smooth mapping $F:R\times\mathcal{P}_1\mapsto
R\times\mathcal{P}_2$ is called a {\it canonical transformation}
if each of the following holds:
\begin{description}
    \item[D1] $F$ is a diffeomorphism;
    \item[D2] $F$ preserves time; that is, $F^*t=t$;
    \item[D3] there is a function
    $K_F\in\mathcal{C}^{\infty}(R\times\mathcal{P}_1)$ such that
    $F^*\tilde{\omega}_2=\omega_K$, where $\omega_K=\tilde{\omega}_1+dK_F\wedge dt$.
\end{description}

In what follows, we will use the symbol $H_T$ instead of $K_F$. There are a few observations that can be made in effort to understand the above definition. First, we note that the slicing of the constraint surface:
\begin{equation}
\mathcal{S}\xrightarrow{~t~} R
\end{equation}
is needed in order to introduce the canonical transformations.
However, the lack of this slicing is the essence of gauge
invariance and it can only be {\it postulated}. Once it is done, {\it any}
canonical transformation, according to condition {\bf D2}, {\it
preserves chosen time}. 

This leads us to the next observation that different choices of time must produce canonically inequivalent theories. It confirms our earlier result that the choice of slicing fixes the Poisson structure, which is related to the defining of an ambiguous procedure by which a non-invertible matrix ($\omega_S$) can be inverted ($\omega_{S,t}^{-1}$).

Another observation is as follows. Suppose there is a given slicing, $t$. The constraint surface can be parametrized as $\mathcal{S}=t\times\mathcal{P}_R$, where $\mathcal{P}_R$ is the reduced phase space. We note that $\omega_{S,t}=\omega_R$ and that $\omega_S=\widetilde{\omega}_R(=t^*\omega_R)$. In this contact manifold there is no Hamiltonian, since the reduced phase space consists of Dirac observables for which $\dot{\mathcal{D}}_i=0$. Now we may introduce the evolution into the system by considering time-dependent reparametrization of the constant time hypersurfaces. As we will see below this will render a non-vanishing Hamiltonian.

Consider the following canonical transformation $F: \mathcal{S}\mapsto\mathcal{S}$, such that $F^*:\mathcal{D}_i\mapsto \mathcal{D}_i(q,p,t)$, $F^*t=t$ and:
\begin{equation}\label{transF}
F^*\omega_S=\widetilde{\omega}_{S,t}+dH_T\wedge dt,
\end{equation}
where $\widetilde{\omega}_{S,t}=t^*{\omega}_{S,t}$ is the pullback of the symplectic from $\omega_{S,t}$ living in the leaf $t=const$, and parametrized with new coordinates $(q,p)$. The new coordinates are in general time-dependent.

In the coordinates $(\mathcal{D}_i,t)$, the null vector of the form $\omega_S$ is given by $X_H=\partial_t$. The canonical transformation $F$ changes the coordinates  and the coordinate expression for $X_H$ accordingly:
\begin{equation}
F^*X_H=\partial_t+X_T=\partial_t-\omega_{S,t}^{-1}(\cdot,H_T)=\partial_t+\{\cdot,H_T\}\bigg|_t
\end{equation}
Now it is seen that the evolution of the observables in new coordinates is not only given through the explicit dependence on time $t$ but also through the true Hamiltonian $H_T$.
 
However, not all time-preserving diffeomorphisms are canonical transformations. Therefore it is useful to introduce an alternative
formulation which relies on a generating function W:
\begin{equation}
p_idq^i+H_Tdt-F^*\theta_R=dW
\end{equation}
where $\theta_{R}$, satisfying $-d\theta_{R}=\omega_R$, is the
canonical (Poincare) one-form in the reduced phase space, and
$(p_i,q^i)$ are the new canonical pairs. $W$ is a generating
function such that\footnote{We denote by ${q}_{D}^i$ and
$p_{D,i}$ the reduced phase
space basic variables.}:
\begin{equation}
W=W({q}_{D}^i,q_i,t),~~{p}_{D,i}=-\frac{\partial
W}{\partial {q}_{D}^i},~~p_i=\frac{\partial W}{\partial q^i}
\end{equation}
and the relation between $W$ and $H_T$ reads:
\begin{equation}
H_T\bigg(t,{q}^i,\frac{\partial W}{\partial
{q}^i}\bigg)-\frac{\partial W}{\partial t}=0
\end{equation}

The final remark is that the true Hamiltonian is quite arbitrary and it does not depend on the particular choice of time but rather on the choice of basic variables on the constant time submanifolds once time is given.

Let us sum up. In the kinematical phase space $\mathcal{P}$, the symplectic form,
$\omega$, and the Poisson bracket, $-\omega^{-1}$, can be
considered interchangeably. However, in the constraint surface
$\mathcal{S}$ the induced two-form, $\omega_S$, is
singular and one cannot define the Poisson bracket in
$\mathcal{S}$ unambiguously. The bracket is needed to compute the
commutation relations between partial observables prior to quantization. The choice of the clock variable
$t:\mathcal{S}\mapsto R$, is
equivalent to the choice of the following commutation relation:
\begin{equation}\label{timepois}
\{t,\mathcal{D}_i\}_S=0
\end{equation}
The last relation fixes the Poisson bracket between {\it any} pair of
partial observables on $\mathcal{S}$. In principle, one can choose {\it any}
function $t:\mathcal{S}\mapsto R$ admitting $X_H(t)>0$ to
slice the constraint manifold. The fact, that different slicings
lead to canonically inequivalent theories is confirmed by the formula (\ref{timepois}).

It should be emphasized that the Poisson bracket between any
pair of Dirac observables is given uniquely in a constrained system. There is the unique two-form in the reduced phase space,
$\omega_R$, such that its pullback to the constraint surface gives
$\pi^*(\omega_R)=\omega_S$. The form $\omega_R$ is invertible,
since the reduced phase space $\mathcal{P}_R$ is even-dimensional,
and the Poisson bracket can be computed. This is in agreement with the fact that the Poisson 
bracket between Dirac observables in the kinematical phase space, 
$\mathcal{P}$, is given uniquely and independently of the choice of time.

\subsection{Problem of time in Dirac quantization}\label{dir}

In the Dirac quantization, the kinematical phase space is
quantized so that the kinematical Hilbert space is obtained and the Hamiltonian constraint is promoted to a self-adjoint
operator $H\mapsto\hat{H}$. Then the theory is constructed
via the solutions to the quantum constraint equation, i.e.
\begin{equation}\label{qcon}
\hat{H}\psi=0
\end{equation}
The solutions $\psi$ normally do not belong to the kinematical
Hilbert space, and the Hilbert space structure needs to be
reintroduced. There is the idea, called
`deparameterization', which cures the problem and at the same time
nicely introduces the concept of evolution of the system. The idea
is to reformulate the constraint equation $H=0$ in such a way that
the quantum constraint equation (\ref{qcon}) gets a
Schr\"odinger-like form \cite{lew3,lew2}. This procedure is very
closely related to another procedure used for the Dirac quantization, namely the group avaraging method \cite{lew}. Therefore, we will focus here only on the idea of deparametrization while keeping in mind that the problem of time is in fact method-independent and can be also formulated in the context of group averaging. 

Suppose that we consider a gravitational system including a scalar field. In this case (see e.g. \cite{lew2}), $H\approx p_{\phi}^2-C_{GR}$ and we obtain a Schr\"odinger-like equation:
\begin{equation}
-i\hbar\frac{d}{d\phi}\psi=\sqrt{\hat{C}}_{GR}\psi
\end{equation}
so that the scalar field, $\phi$, plays a
role of a time parameter. The non-vanishing true Hamiltonian, $\sqrt{\hat{C}}_{GR}$, is
expressed in terms of the rest of the kinematical degrees of
freedom, here the gravitational ones. They are supposed to
parameterize the reduced phase space and play a role of (partial)
observables in quantum theory. It must be noted, however, that the
commutation relations between these partial observables are {\it
postulated}. They are parachuted from the kinematical phase space
in an {\it ad hoc} manner. Since the observables are
physical only in the constraint surface, any kinematical degree of
freedom, $\mathcal{P}_i$, forms the following equivalence class:
\begin{equation}
\mathcal{P}_i\sim\mathcal{P}'_i~\Longleftrightarrow~\mathcal{P}_i
\approx\mathcal{P}'_i~\Longleftrightarrow~\mathcal{P}_i=\mathcal{P}'_i+C_i
\end{equation}
where $C_i$ is a constraint. But then the Poisson bracket between
the equivalence classes on the constraint surface is quite easily shown to be ill-defined (non-unique):
\begin{equation}\label{ec}
\{\mathcal{P}_i+C_i,\mathcal{P}_j+C_j\}=\{\mathcal{P}_i,\mathcal{P}_j\}+\{C_i,\mathcal{P}_j\}
+\{\mathcal{P}_i,C_j\}+\{C_i,C_j\}
\end{equation}
where only the term $\{C_i,C_j\}\approx 0$ weakly vanishes. The
terms $\{C_i,\mathcal{P}_j\}$ and $\{\mathcal{P}_i,C_j\}$ do not
vanish in the constraint surface and make the Poisson structure ill-defined.

The non-existence of the Poisson bracket in the constraint surface
$\mathcal{S}$ was proved in the previous subsection and is due to
the fact that the constraint surface is odd-dimensional.
Therefore, in order to encode dynamics into quantum theory, one
{\it needs} to postulate the Poisson bracket between partial observables or between a partial and Dirac one, so that relation (\ref{timepois}) is fixed. Having this done,
the time parameter is determined as the {\it only} partial
observable which commutes with all the other, partial and Dirac,
observables. Then, the time parameter $t:\mathcal{S}\mapsto R$,
together with the Dirac observables, introduces the `no-Hamiltonian' parameterization
of the constraint surface $\mathcal{S}= R\times\mathcal{P}_R$. In the scalar field case $t=\phi$, one has:
\begin{equation}
\{\phi,\mathcal{D}_i\}_S=0
\end{equation}

Observe once again that in a constrained system, only the Poisson commutation
between any pair of Dirac observables whatever its parametrization (see Appendix \ref{dirac}) is well-defined:
\begin{equation}\label{poisson}
\{\mathcal{D}_i+C_i,\mathcal{D}_j+C_j\}=\{\mathcal{D}_i,\mathcal{D}_j\}+\{\mathcal{D}_i,C_j\}+\{C_i,\mathcal{D}_j\}
+\{C_i,C_j\}\approx \{\mathcal{D}_i,\mathcal{D}_j\}
\end{equation}
since a Dirac observable by definition commutes weakly with a
constraint, i.e. $\{\mathcal{D}_i,C_j\}\approx 0$ (in opposition
to what happens in (\ref{ec})).

To sum up, in the Dirac quantization, the problem of time persists. As before, this is so due to the fact that time-dependent observables are gauge-variant quantities and their Poisson commutation relations are undefined. In this procedure, an ambiguous Poisson structure can be postulated by parachuting the Poisson structure of $2n-2$ kinematical degrees of freedom of ambiguous choice.

\section{Choice of time and spectra of partial observables}
So far we have showed that the Poisson bracket between partial observables depends on the choice of time. In what follows, we will show how this fact affects spectral properties of time-dependent quantities.
Let us study the Hubble observable in a fixed direction (\ref{h4}):
\begin{equation}\label{h1fix}
H_1= \frac{1-\cos\phi+2\sin\phi}{\sqrt{2}(7+\cos\phi)~t_{cos}}
\end{equation}
First we notice that the evolution of the Hubble observable, in the cosmological time $t_{cos}$, is not canonical. For from (\ref{h1fix}), we have:
\begin{equation}
d\phi=\frac{\sqrt{2}t_{cos}(7+\cos\phi)^2}{2+8\sin\phi+14\cos\phi}(dH_1+\frac{H_1}{t_{cos}}dt_{cos})
\end{equation}
which after substituting for $d\phi$ in (\ref{romega}) leads to
\begin{equation}
\omega_S=\dots+\frac{\sqrt{2}H_1(7+\cos\phi)^2}{2+8\sin\phi+14\cos\phi}~d\mathcal{O}_2\wedge dt_{cos} 
\end{equation}
which according to the theory of canonical transformations (see the definition in the previous section) should have the form of (\ref{transF}), so that
the one-form\begin{equation}\frac{\sqrt{2}H_1(7+\cos\phi)^2}{2+8\sin\phi+14\cos\phi}~d\mathcal{O}_2\end{equation} should be equal to the derivative of a true Hamiltonian, $dH_T$, which here would be a generator of the canonical motion of $H_1$. {\it The existence of such a generator would allow us to construct a quantum theory in which the evolution of $H_1$ would be unitary}. This apparently does not hold, since the above one-form is not closed. In what follows, we will replace the cosmological time with a different clock variable.

For brevity, let us restrict to $\phi\in (0, \phi_1)$\footnote{This phase space sector consists of all the solutions modulo the interchange of the axes in the homogenous leaf.} and redefine time $t_{cos}$ in two ways:
\begin{eqnarray}\label{t1}
t_{cos}&=&\frac{1-\cos\phi+2\sin\phi}{\sqrt{2}(7+\cos\phi)}(t_{1}+O_1)\\ \label{t2}
t_{cos}&=&\frac{1-\cos\phi+2\sin\phi}{\sqrt{2}(7+\cos\phi)}\bigg(\frac{\pi}{\phi_1}\sin(\frac{\pi}{\phi_1}\phi)t_{2}+O_2\bigg)
\end{eqnarray}
Both $t_1$ and $t_2$ are well-defined, with $t_{1}\in(-O_1,\infty)$ for a given $O_1$ and $t_{2}\in(-\frac{O_1}{\frac{\pi}{\phi_1}\sin(\frac{\pi}{\phi_1}\phi)},\infty)$ for a given $O_1$ and $\phi$. The times are half-lines and the left endpoints signal the singularity. Now, the Hubble observable (\ref{h1fix}) takes the form:
\begin{equation}
H_1=\frac{1}{t_{1}+O_1}=\frac{1}{\frac{\pi}{\phi_1}\sin(\frac{\pi}{\phi_1}\phi)t_{2}+O_2}
\end{equation}
The difference between the above formulae is due to the different choice of time, i.e. a different parametrization of the constraint surface, though the function $H_1:\mathcal{S}\mapsto R$ itself remains unchanged, i.e. attaches real numbers to points in the constraint surface uniquely.

Let us see that definitions (\ref{t1}) and (\ref{t2}) lead to canonical motion in $H_1$.  First let us examine the evolution in $t_1$. Introduce:
\begin{equation}\label{T1}
{T}_1:=t_1+\mathcal{O}_1
\end{equation}
so that
\begin{equation}\label{T12}
{H}_1=\frac{1}{{T}_1}
\end{equation}
and
\begin{equation}\label{T13}
\omega_S=dT_1\wedge dr+d\mathcal{O}_2\wedge d\phi +dH_T\wedge dt_1
\end{equation}
where $H_T=r$ is a true Hamiltonian, which generates the canonical motion of $H_1$, in $t_1$. For each constant time slice the symplectic form (\ref{T13}) may be inverted to the Poisson structure:
\begin{equation}\label{T14}
\{\cdot ,\cdot\}\bigg|_{t_1=const}=\frac{\partial~\cdot}{\partial T_1}\frac{\partial~\cdot}{\partial r}-\frac{\partial~\cdot}{\partial r}\frac{\partial~\cdot}{\partial T_1}+\frac{\partial~\cdot}{\partial \mathcal{O}_2}\frac{\partial~\cdot}{\partial \phi}-\frac{\partial~\cdot}{\partial \phi}\frac{\partial~\cdot}{\partial \mathcal{O}_2}
\end{equation}
Now let us examine the evolution in $t_2$. Introduce:
\begin{equation}\label{T2}
{T}_2:=\frac{\pi}{\phi_1}\sin(\frac{\pi}{\phi_1}\phi)t_{2}+O_2
\end{equation}
so that
\begin{equation}\label{T22}
{H}_1=\frac{1}{{T}_2}
\end{equation}
and
\begin{equation}\label{T23}
\omega_S=d\mathcal{O}_1\wedge dr+dT_2\wedge d\phi +dH_T\wedge dt_2
\end{equation}
where $H_T=-\cos(\frac{\pi}{\phi_1}\phi)$ is a true Hamiltonian, which generates the canonical motion of $H_1$, in $t_2$. For each constant time slice the symplectic form (\ref{T23}) may be inverted to the Poisson structure:
\begin{equation}\label{T24}
\{\cdot ,\cdot\}\bigg|_{t_2=const}=\frac{\partial~\cdot}{\partial \mathcal{O}_1}\frac{\partial~\cdot}{\partial r}-\frac{\partial~\cdot}{\partial r}\frac{\partial~\cdot}{\partial \mathcal{O}_1}+\frac{\partial~\cdot}{\partial T_2}\frac{\partial~\cdot}{\partial \phi}-\frac{\partial~\cdot}{\partial \phi}\frac{\partial~\cdot}{\partial T_2}
\end{equation}

\subsection{Warm-up: Quantization of Dirac observables}
We have identified the four-dimensional space of Dirac
observables, $(r,\phi,\mathcal{O}_1,\mathcal{O}_2)\in R_+\times
(0,\phi_1)\times R\times R$, equipped with the
symplectic form $\omega_R$ identical with
$\omega_S$ in (\ref{romega}). By inverting the form in the reduced phase space one can find
the Poisson bracket:
\begin{equation}
\{\mathcal{O}_1,r\}=1,~~\{\mathcal{O}_2,\phi\}=1
\end{equation}
We assign the following operators to the corresponding Dirac
observables:
\begin{equation}
r\mapsto
\hat{r}:=r,~~\mathcal{O}_1\mapsto\hat{\mathcal{O}}_1:=i\partial_{r},~~
\phi\mapsto
\hat{\phi}:=\phi,~~\mathcal{O}_2\mapsto\hat{\mathcal{O}}_2:=i\partial_{\phi}
\end{equation}
so that the following algebra homomorphism is obtained:
\begin{equation}
\widehat{\{\mathcal{O}_1,r\}}=\frac{1}{i}[\hat{\mathcal{O}}_1,\hat{r}]
,~~\widehat{\{\mathcal{O}_2,\phi\}}=\frac{1}{i}[\hat{\mathcal{O}}_2,\hat{\phi}]
\end{equation}
and the physical Hilbert space is defined as:
\begin{equation}
\mathcal{H}_{phys}=L^2(R_+\times (0,\phi_1),drd\phi)
\end{equation}
The formally self-adjoint operator $i\partial_{r}$, defined
on a half-line, can be made into a self-adjoint operator by the
following assignment:
\begin{equation}
i\partial_{r}:=\sqrt{-\partial_{r}^2}
\end{equation}
where the Laplacian $-\partial_{r}^2$ acts on the closure
(in the `operator norm') of the following space
\cite{Der}:
\begin{equation}\label{domht}
\{\psi\in
\mathcal{H}_{phys}:~\psi(r=0)=\mu\cdot\partial_r\psi(r=0)\}
\end{equation}
where $\mu\geq 0$ enumerates unitarily inequivalent self-adjoint
realizations of $-\partial_{r}^2$. The values $\mu=0$ and
$\mu=\infty$ correspond to the Dirichlet and Neumann condition,
respectively. The spectrum reads:
\begin{equation}
sp_{R_+}\{i\partial_{r}\}=R_+
\end{equation}

The formally self-adjoint operator $i\partial_{\phi}$ is also
unbounded and enjoys many unitarily inequivalent essentially self-adjoint
realizations in the space \cite{Der}:
\begin{equation}
\{\psi\in
\mathcal{H}_{phys}:~\psi(0)=\psi(\phi_1)e^{i2\pi\kappa}\}
\end{equation}
where $0$ and $\phi_1$ are the boundary points of the
domain. The parameter $\kappa\in [0,1)$
enumerates unitarily inequvalent representations of
$i\partial_{\phi}$. Its spectrum reads
\begin{equation}
sp\{i\partial_{\phi}\}=\frac{\kappa}{\phi_1}+\frac{2\pi}{\phi_1} n,~~n\in\mathbb{Z}
\end{equation}

This completes the quantization of all the gauge-invariant
observables in the Kasner universe. Since the notion of evolution
is absent in this model, one cannot ask questions about the fate
of singularity.

\subsection{Quantization of geometrical observables in two different time variables}
\subsubsection{Case $t_1$}
Let us use the Schr\"odinger representation to study the
Hubble observable defined in (\ref{T12}) with the Poisson structure given in (\ref{T14}):
\begin{equation}
r\mapsto
\hat{r}:=-i\frac{d}{dr},~~{T}_{1}\mapsto\hat{T}_{1}:=r,~~\phi\mapsto\hat{\phi}:=\phi,
~~\mathcal{O}_2\mapsto\hat{\mathcal{O}}_2:=-i\frac{d}{d\phi}
\end{equation}
which act on the Hilbert space
\begin{equation}
\mathcal{H}_{phys}=L^2(R_+\times (0,\phi_1),drd\phi)
\end{equation}
so that the following is satisfied:
\begin{equation}
\widehat{\{T_{1},r\}}=\frac{1}{i}[\hat{T}_{1},\hat{r}]
,~~\widehat{\{\mathcal{O}_2,\phi\}}=\frac{1}{i}[\hat{\mathcal{O}}_2,\hat{\phi}]
\end{equation}
The true Hamiltonian reads:
\begin{equation}
\hat{H}_T=-i\frac{d}{dr}
\end{equation}
The operator $-i\frac{d}{dr}$ was
discussed in the previous subsection, and we showed that it is
essentially self-adjoint in the domain (\ref{domht}). Thus, due to
the Stone-von Neumann theorem there exists a unitary operator
\begin{equation}
U=e^{-it\hat{H}_T}
\end{equation}
which proves that the dynamics of the system is well defined for
{\it all} $t\in R$. This means that {\it the singularity which is reached in finite
time in classical theory is resolved at quantum level}. Note that
the quantum operators associated with physical measurements (like connection or curvature) may be {\it unbounded}. These circumstances, however,
are common in quantum theory and do not spoil the singularity
resolution.

Let us move to the quantization of $H_1$:
\begin{equation}
\hat{H}_1=\frac{1}{\hat{T}_1}=\frac{1}{r},~~r\in R_+
\end{equation}
which is a well-defined self-adjoint (unbounded) operator on a half-line with the {\it continuous} spectrum $sp_{t_1}(H_1)=R_+$. 

\subsubsection{Case $t_2$}
Let us use the Schr\"odinger representation to study the
Hubble observable defined in (\ref{T22}) with the Poisson structure given in (\ref{T24}):
\begin{equation}
r\mapsto
\hat{r}:=-i\frac{d}{dr},~~\mathcal{O}_{1}\mapsto\hat{\mathcal{O}}_{1}:=r,~~\phi\mapsto\hat{\phi}:=\phi,
~~{T}_2\mapsto\hat{T}_2:=-i\frac{d}{d\phi}
\end{equation}
which act on the Hilbert space
\begin{equation}
\mathcal{H}_{phys}=L^2(R_+\times (0,\phi_1),drd\phi)
\end{equation}
so that the following is satisfied:
\begin{equation}
\widehat{\{\mathcal{O}_{1},r\}}=\frac{1}{i}[\hat{\mathcal{O}}_{1},\hat{r}]
,~~\widehat{\{T_2,\phi\}}=\frac{1}{i}[\hat{T}_2,\hat{\phi}]
\end{equation}
The true Hamiltonian reads:
\begin{equation}
\hat{H}_T=-\cos\big(\frac{\pi}{\phi_1}\phi\big)
\end{equation}
The operator $-\cos(\frac{\pi}{\phi_1}\phi)$ is bounded, symmetric and hence self-adjoint in the Hilbert space. Due to
the Stone-von Neumann theorem there exists a unitary operator
\begin{equation}
U=e^{-it\hat{H}_T}
\end{equation}
which proves that the dynamics of the system is well defined for
{\it all} $t\in R$. This again means that {\it the singularity which is reached in finite time in classical theory is resolved at quantum level}.

Let us move to the quantization of $H_1$:
\begin{equation}
\hat{H}_1=\frac{1}{\hat{T}_2}
\end{equation}
The operator:
\begin{equation}
\hat{T}_2=-i\frac{d}{d\phi}
\end{equation}
has been already discussed and is a well-defined self-adjoint (unbounded) operator with the discrete spectrum $sp\{i\partial_{\phi}\}=\frac{\kappa}{\phi_1}+\frac{2\pi}{\phi_1} n$. Using the spectral theorem and requiring positivity of $\hat{H}_1$, we find that the spectrum of the Hubble observable is {\it discrete}, {\it bounded} and reads $sp_{t_2}(\hat{H}_1)=\bigg\{ \bigg|\frac{1}{\frac{\kappa}{\phi_1}+\frac{2\pi}{\phi_1} n}\bigg|,~~n\in\mathbb{Z}\bigg\}$..

\section{Conclusions}

Motivated by the BKL scenario, we started this paper aiming at deriving a quantum theory of the Kasner epoch. This task seemed to be feasible with the use of the reduced phase space method: we defined the kinematical phase space, in which we solved Hamilton's equation of motion; then we identified Dirac observables and their algebra; finally, we arrived at the physical solutions of the classical theory in terms of Dirac observables and a clock variable. At this point, the quantization is usually performed. However, we realized that the choice of clock variable determined the functional dependance of time-dependent quantities on Dirac observables. In the rest of the paper, we studied the consequences of this fact.

We showed that in addition to the usual
ambiguities of quantum theory, in a Hamiltonian constraint
system like general relativity, there is also another ambiguity related to
the choice of the clock. We have managed to clarify the procedure of
encoding evolution in quantum gravity. It turned out that the
procedure could be identified with inverting a singular matrix. The
matrix is the induced two-form on the constraint surface
$\omega_S$ and its inverse, in a sense, still exists but is
ambiguous and represents the Poisson bracket. Different choices
of the inverse lead to canonically inequivalent classical
theories, which are then quantized. Then the dependance of quantum physics on the choice of time variable can be stated as follows: The Poisson bracket associates a canonical transformation with each phase space function. In classical theory, the Poisson bracket is an auxiliary mathematical structure, which does not affect the physical (observable) content of the theory in the sense that it does not matter which canonical transformation corresponds to which phase space function as long as the equations of motion are equivalent. In quantum theory, however, the physical quantities are associated with operators on the Hilbert space, in which they act as generators of unitary transformations. The correspondence between classical and quantum theory is the one between the canonical transformations and the unitary ones. Therefore, the Poisson structure is not auxiliary but essential for this correspondence and thus all ambiguities in canonical formulations of classical theory are expected to lead to quantum theories with different {\it physical} content. Thus, for each
choice of time, one may find a distinct spectrum for a given
partial observable. We have shown that this is the case for a directional Hubble parameter for which we were able to obtain both continuous and discrete spectrum, depending on the choice of the clock variable.
Which spectrum, if any, is the correct one? Answering this question is beyond the framework of general relativity
and usual quantum mechanics.

In the view of this result, one should construct all the quantum
theories treating all the possible time parameters on equal
footing. The relation between the resultant canonically
inequivalent theories surely deserves further study. This will be
the subject of next papers by the present author.

What is the physical interpretation of the obtained result
that theories of gravity with different time
parameters are canonically inequivalent? Is the evolution of the
universe not an objective reality but merely an {\it impression}
perceived by an observer? Is it possible for every observer to
`measure' his own story of the big bang? At this moment, it is very tempting to speculate that perhaps {\it
the fundamental quantum theory we all are looking for is not a
gauge theory - the gauge invariance is obtained only in the
classical limit, in which the uncertainty principle is removed}.
This point of view is strongly supported by the fact that one can
quite easily reformulate a classical non-relativistic system in a
gauge-invariant manner (see e.g. \cite{rovel}). The natural
question then arises: among all the partial observables,
which one is {\it always} classical?

\begin{acknowledgments}
This work was supported by the Foundation for Polish Science fellowship KOLUMB. I would like to thank Professors W\l odzimierz Piechocki, Abhay Ashtekar and Martin Bojowald for helpful
discussions.
\end{acknowledgments}

\appendix
\section{Calculation of connection and curvature}\label{calc} We consider a manifold $\mathcal{M}$
equipped with the metric:
\begin{equation}
ds^2=-N^2dt^2+\sum_ia_i^2(dx^i)^2
\end{equation}
where $i=1,2,3$, and on which we introduce the vector fields and
the dual 1-forms:
\begin{equation}
e_0=\frac{1}{N}\partial_t~,~~e_i=\frac{1}{a_i}\partial_i,~~\sigma^0=Ndt~,~~\sigma^i=a_idx^i
\end{equation}
so that the following relations hold:
\begin{equation}
e_{\mu}\cdot
e_{\nu}=\eta_{\mu\nu}~,~~\sigma^{\mu}(e_{\nu})=\delta^{\mu}_{\nu}
\end{equation}
where $\mu,\nu=0,1,2,3$. We assume that \cite{Fl}:
\begin{equation}
d\big(\sum_{\mu}\sigma^{\mu}e_{\mu}\big)=\sum_{\mu}\big(d\sigma^{\mu}-\sum_{\nu}\sigma^{\nu}\omega_{~\nu}^{\mu}\big)e_{\mu}=0
\end{equation}
and obtain the connection $\Omega$:
\begin{equation}
d\sigma=\sigma\Omega~,~~\Omega=\|\omega_{~\nu}^{\mu}\|
\end{equation}
which is a matrix of 1-forms. It is uniquely determined from the
condition $\omega_{\nu\mu}+\omega_{\mu\nu}=0$, and equals:
\begin{equation}
\Omega=\left(%
\begin{array}{cccc}
  0 & \frac{\dot{a}_1}{Na_1}\sigma^1 & \frac{\dot{a}_2}{Na_2}\sigma^2 & \frac{\dot{a}_3}{Na_3}\sigma^3 \\
  \frac{\dot{a}_1}{Na_1}\sigma^1 & 0 & 0 & 0 \\
  \frac{\dot{a}_2}{Na_2}\sigma^2 & 0 & 0 & 0 \\
  \frac{\dot{a}_3}{Na_3}\sigma^3 & 0 & 0 & 0 \\
\end{array}%
\right)
\end{equation}
From $\Omega$ we compute the curvature matrix $\Theta$:
\begin{equation}
\Theta=\|\theta_{~\nu}^{\mu}\|=d\Omega-\Omega^2
\end{equation}
which is equal to
\begin{equation}
\Theta=\left(%
\begin{array}{cccc}
  0 & \frac{1}{Na_1}\big(\frac{\dot{a}_1}{N}\big)_{,t}\sigma^0\sigma^1 & \frac{1}{Na_2}\big(\frac{\dot{a}_2}{N}\big)_{,t}\sigma^0\sigma^2 & \frac{1}{Na_3}\big(\frac{\dot{a}_3}{N}\big)_{,t}\sigma^0\sigma^3 \\
  \frac{1}{Na_1}\big(\frac{\dot{a}_1}{N}\big)_{,t}\sigma^0\sigma^1 & 0 & -\big(\frac{\dot{a}_1}{Na_1}\big)\big(\frac{\dot{a}_2}{Na_2}\big)\sigma^1\sigma^2 & -\big(\frac{\dot{a}_1}{Na_1}\big)\big(\frac{\dot{a}_3}{Na_3}\big)\sigma^1\sigma^3 \\
  \frac{1}{Na_2}\big(\frac{\dot{a}_2}{N}\big)_{,t}\sigma^0\sigma^2 & \big(\frac{\dot{a}_1}{Na_1}\big)\big(\frac{\dot{a}_2}{Na_2}\big)\sigma^1\sigma^2 & 0 & -\big(\frac{\dot{a}_2}{Na_2}\big)\big(\frac{\dot{a}_3}{Na_3}\big)\sigma^2\sigma^3 \\
  \frac{1}{Na_3}\big(\frac{\dot{a}_3}{N}\big)_{,t}\sigma^0\sigma^3 & \big(\frac{\dot{a}_1}{Na_1}\big)\big(\frac{\dot{a}_3}{Na_3}\big)\sigma^1\sigma^3 & \big(\frac{\dot{a}_2}{Na_2}\big)\big(\frac{\dot{a}_3}{Na_3}\big)\sigma^2\sigma^3 & 0 \\
\end{array}%
\right)
\end{equation}
The Riemann curvature tensor is now given by:
\begin{equation}
\theta_{~\nu}^{\mu}=\frac{1}{2}\sum_{\alpha,
\beta}R_{~\nu\alpha\beta}^{\mu}\sigma^{\alpha}\sigma^{\beta}
\end{equation}
from which we calculate the Ricci tensor $R_{\nu\beta}$:
\begin{equation}
\|R_{\nu\beta}\|=\left(%
\begin{array}{cccc}
  R_{00} & 0 & 0 & 0 \\
  0 & R_{11} & 0 & 0 \\
  0 & 0 & R_{22} & 0 \\
  0 & 0 & 0 & R_{33} \\
\end{array}
\right)
\end{equation}
where
$R_{00}=-\sum_i\frac{1}{Na_i}\big(\frac{\dot{a}_i}{N}\big)_{,t}$
and
$R_{ii}=\frac{1}{Na_i}\big(\frac{\dot{a}_i}{N}\big)_{,t}\sum_{k\neq
i}\big(\frac{\dot{a}_i}{Na_i}\big)\big(\frac{\dot{a}_k}{Na_k}\big)$.
Thus, the Ricci scalar:
\begin{equation}
R=2\bigg[\sum_i\frac{1}{Na_i}\bigg(\frac{\dot{a}_i}{N}\bigg)_{,t}+\sum_{i>j}\bigg(\frac{\dot{a}_i}{Na_i}\bigg)\bigg(\frac{\dot{a}_j}{Na_j}\bigg)\bigg]
\end{equation}
In the Friedman cosmology, it is common to use the Hubble parameter
$H$ and the deceleration parameter $q$, which in the case of
Kasner universe can also be introduced, for each of the three
directions separately:
\begin{equation}
H_i=\frac{\dot{a}_i}{Na_i}~,~~q_i=-\frac{\frac{1}{Na_i}\big(\frac{\dot{a}_i}{N}\big)_{,t}}{(\frac{\dot{a}_i}{Na_i})^2}=-\frac{\frac{1}{N}\dot{H}_i+H_i^2}{H_i^2}
\end{equation}
Note that these parameters determine all the components of the
curvature matrix $\Theta$.

\section{Parametrization of Dirac observables}\label{dirac}
It is important to realize the precise notion of Dirac
observable. It is sometimes claimed that Dirac observable is a
kinematical phase space function, which commutes weakly with the
Hamiltonian constraint. This is not precise enough. In a constrained
system, the physical motion is realized in the constraint surface,
let us say $H=0$. This equality implicates the embedding of the
constraint surface into the kinematical phase space,
$E:\mathcal{S}\mapsto\mathcal{P}$. One cannot pull back functions
from $\mathcal{S}$ to functions on $\mathcal{P}$, since the range
of $E$ is restricted only to $H=0$. This means that the Dirac
observable cannot be defined as a function of the kinematical
phase coordinates. The proper definition is as follows: Dirac's
observable $\mathcal{D}_i$ is a constraint surface function, which
commutes with the Hamiltonian constraint $H$, i.e.
$X_{H}(\mathcal{D}_i)\big|_S=0$, where $X_{H}$ satisfies
$\omega(\cdot,X_{H})=dH$.

Let us find functions $\mathcal{O}_D$, which satisfy:
\begin{equation}\label{dirprec}
\{\mathcal{O}_D, H\}= 0
\end{equation}
In the Kasner universe the above equality in the convenient gauge
$N=4\sqrt{2}\sqrt{X_1X_2X_3}$ takes the following form:
\begin{equation}
\sum_{i\neq j\neq
k}(X_jP_j+X_kP_k)\bigg(P^i\frac{\partial\mathcal{O}_D}{\partial
P^i}-X_i\frac{\partial\mathcal{O}_D}{\partial X_i}\bigg)=0
\end{equation}
and we obtain:
\begin{equation}
\mathcal{O}_D=\mathcal{O}_D(\Gamma_i, \Omega_{ij})
\end{equation}
where
\begin{equation}
\Omega_{ij}=(\Gamma_i+\Gamma_j)(\Gamma_i+\Gamma_k)\ln|P_i|-(\Gamma_i+\Gamma_j)(\Gamma_j+\Gamma_k)\ln|P_j|
\end{equation}
and satisfy the relation:
\begin{equation}
\Omega_{ij}-\Omega_{kj}=\Omega_{ik}
\end{equation}
These are solutions to (\ref{dirprec}) in a given gauge. Their restriction to the constraint surface
(\ref{con3}) is gauge-invariant and gives the four-dimensional space of Dirac
observables, $\mathcal{P}_R$. We could alternatively define the Dirac
observable as equivalence classes with the equivalence relation:
\begin{equation}
\mathcal{O}_D\sim\mathcal{O}'_D~\Longleftrightarrow~\mathcal{O}_D\approx\mathcal{O}'_D~\Longleftrightarrow~\mathcal{O}_D=\mathcal{O}'_D+C
\end{equation}
where `$\approx$' denotes `equals on the constraint surface' and
$C\approx 0$ is a constraint. This, however, leads to the
ambiguity in `parameterization' of Dirac observable as there is no
natural projection from the kinematical phase space to the reduced phase
space. The important issue here is that the Poisson bracket is
well defined between the equivalence classes (see equation
(\ref{poisson})).

\section{The Kasner metric in terms of the cosmological time and Dirac observables}
Let us express the physical solutions in terms of Dirac
observables and cosmological time $t_{cos}$:
\begin{eqnarray}\nonumber
ds^2=-dt_{cos}^2&+&2\tilde{a}_1^2\bigg(-\frac{r(7+\cos\phi)}{16\sqrt{2}}~t_{cos}\bigg)^{2\cdot\frac{2-2\cos\phi+4\sin\phi}{7+\cos\phi}}(dx^1)^2\\
&+&2\tilde{a}_2^2\bigg(-\frac{r(7+\cos\phi)}{16\sqrt{2}}~t_{cos}\bigg)^{2\cdot\frac{2-2\cos\phi-4\sin\phi}{7+\cos\phi}}(dx^2)^2\\
\nonumber
&+&2\tilde{a}_3^2\bigg(-\frac{r(7+\cos\phi)}{16\sqrt{2}}~t_{cos}\bigg)^{2\cdot\frac{3+5\cos\phi}{7+\cos\phi}}(dx^3)^2
\end{eqnarray}
where
\begin{eqnarray}\nonumber
\tilde{a}_1&=&
e^{\frac{\{\mathcal{O}_1(7\sin\phi-4\cos\phi)+\frac{1}{r}\mathcal{O}_2(1+3\cos\phi+4\sin\phi)\}\{-5-9\cos\phi-4\sin\phi\}}{2(7+\cos\phi)(5+3\cos\phi-4\sin\phi)}}\\
\tilde{a}_2&=&
e^{\frac{\mathcal{O}_1(\frac{35}{2}\sin\phi+10\cos\phi+\frac{5}{2}\sin\phi\cos\phi-14\sin^2\phi+6\cos^2\phi)+\frac{1}{r}\mathcal{O}_2(15-12\sin\phi+\cos\phi-12\sin\phi\cos\phi-\frac{9}{2}\sin^2\phi)}{(7+\cos\phi)(5+3\cos\phi-4\sin\phi)}}\\
\nonumber\tilde{a}_3&=&
e^{\frac{\mathcal{O}_1(-20\cos\phi+16\sin\phi\cos\phi-12\cos^2\phi)+\frac{1}{r}\mathcal{O}_2(-16+20\sin\phi+8\cos\phi+12\sin\phi\cos\phi+16\cos^2\phi)}{(7+\cos\phi)(5+3\cos\phi-4\sin\phi)}}
\end{eqnarray}
The minus sign in front of the time, $-t_{cos}$, comes from our
convention that as time grows the universe approaches the
singularity, that is $t_{cos}\in (-\infty,0)$. Now, we can relate
the parameters $p_1$, $p_3$ and $p_3$ occurring in the metric
(\ref{taub}) to the Dirac observables:
\begin{eqnarray}\nonumber
p_1&=&\frac{\Gamma_1}{\sum_i\Gamma_i}=\frac{2-2\cos\phi+4\sin\phi}{7+\cos\phi}\\
p_2&=&\frac{\Gamma_2}{\sum_i\Gamma_i}=\frac{2-2\cos\phi-4\sin\phi}{7+\cos\phi}\\
\nonumber
p_3&=&\frac{\Gamma_3}{\sum_i\Gamma_i}=\frac{3+5\cos\phi}{7+\cos\phi}
\end{eqnarray}

We note that all the Dirac observables play a role provided that
the topology of the universe is compact. If, for instance, we set
$\Sigma$ infinite in all directions, then $\phi$ is the only Dirac
observable, since the values of the scale factors ${a}_i$ are
non-physical, and consequently $r$, $\mathcal{O}_1$ and
$\mathcal{O}_2$ are not Dirac observables any longer.

\end{document}